\documentclass[%
 reprint,
showpacs,preprintnumbers,
 amsmath,amssymb,
 aps,
]{revtex4-1}

\usepackage{graphicx}
\usepackage{dcolumn}
\usepackage{bm}

\begin{document}

\title{Cluster nonequilibrium relaxation in Ising models observed with the Binder ratio}

\author{Yoshihiko Nonomura}
\email{nonomura.yoshihiko@nims.go.jp}
\affiliation{International Center for Materials Nanoarchitectonics, 
National Institute for Materials Science, Tsukuba, Ibaraki 305-0044, Japan} 

\author{Yusuke Tomita}
\email{ytomita@shibaura-it.ac.jp}
\affiliation{College of Engineering, Shibaura Institute of Technology, 
Saitama 337-8570, Japan} 

\begin{abstract}
The Binder ratios exhibit discrepancy from the Gaussian behavior of the 
magnetic cumulants, and their size independence at the critical point 
has been widely utilized in numerical studies of critical phenomena. 
In the present article we reformulate the nonequilibrium relaxation 
(NER) analysis in cluster algorithms using the $(2,1)$-Binder ratio, 
and apply this scheme to the two- and three-dimensional Ising models. 
Although the stretched-exponential relaxation behavior at the critical 
point is not explicitly observed in this quantity, we find that there exists 
a logarithmic finite-size scaling formula which can be related with a 
similar formula recently derived in cluster NER of the correlation length, 
and that the formula enables precise evaluation of the critical point 
and the stretched-exponential relaxation exponent $\sigma$. 
Physical background of this novel behavior is explained by 
the simulation-time dependence of the distribution function of 
magnetization in two dimensions and temperature dependence 
of $\sigma$ obtained from magnetization in three dimensions.
\end{abstract}


\maketitle

\section{Introduction}
Finite-size corrections in physical quantities have been 
serious obstacles in numerical studies of critical phenomena. 
They result in poor scaling behaviors, and consequently in poor 
estimates of the critical point and critical exponents. Although a 
straightforward solution is to take correction terms into account, 
increase of fitting parameters also results in poor numerical estimation. 
Then, the Binder ratios~\cite{Binder} defined as the homogeneous 
ratios of (magnetic) cumulants have been widely utilized 
in numerical studies of critical phenomena, because finite-size 
corrections in homogeneous ratios of cumulants of the 
same physical quantities are expected to be cancelled. 
Even though such cancellation may not be perfect, 
correction terms would start from much higher orders. 

Recently we investigated~\cite{Nonomura14,Nonomura15,Nonomura16} 
early-time nonequilibrium relaxation (NER)~\cite{NERrev} 
in cluster algorithms~\cite{SW,Wolff} numerically, and found that 
physical quantities show the stretched-exponential relaxation 
at the critical point, not the power-law one~\cite{NERrev} 
commonly observed in local-update algorithms. 
Although the fusion of cluster algorithms and NER resulted 
in much easier treatment of larger systems than conventional 
equilibrium simulations, resulting estimates of the critical point 
and critical exponents were comparable~\cite{Nonomura16} to 
previous studies in one or two decades ago with much smaller 
systems. The main reason is that we only know the empirical 
scaling form up to the leading order and finite-size corrections 
affect more seriously than in equilibrium simulations. 

Then, we expect that the reformulation of the cluster NER 
using the Binder ratios would improve such a situation, 
while absence of the explicit size dependence in the 
Binder ratios in equilibrium may result in the absence 
of the stretched-exponential relaxation behavior. In the 
present article we analyze the $(2,1)$-Binder ratio defined by 
$B_{2,1}\equiv \langle m^{2}\rangle / \langle |m| \rangle^{2}$, 
which behaves similarly to the commonly-used $(4,2)$-Binder 
ratio~\cite{Binder}. Since step-by-step oscillation between 
positive and negative values of magnetization is observed 
in cluster algorithms, introduction of $B_{2,1}$ including 
$\langle |m| \rangle$ is natural in the present case. 
Quite recently, the present authors derived~\cite{Tomita17} 
a phenomenological finite-size scaling formula of the 
size-independent quantity $\xi/L$ with the correlation length 
$\xi$, and a similar formula is also expected in $B_{2,1}$. 
Furthermore, we also observe simulation-time dependence 
of the distribution function of $|m|$ in order to clarify 
the origin of the stretched-exponential relaxation. 

The outline of the present article is as follows. In section II, 
basic procedures of numerical calculations are summarized, which 
includes the finite-size scaling formula of $\xi/L$ mentioned above. 
In section III, detailed procedures of numerical calculations are 
exhibited through the analysis of the simulation-time dependence 
of the $(2,1)$-Binder ratio in the two-dimensional Ising model 
on a square lattice. We show that the above-mentioned 
finite-size scaling formula of $\xi/L$ also holds in $B_{2,1}$. 
In section IV, simulation-time dependence of the distribution 
function of magnetization is observed, and the origin of the 
stretched-exponential behavior of the critical cluster NER is 
clarified. In section V, similar analysis in the three-dimensional 
Ising model on a simple cubic lattice is described. In section VI, 
the relaxation exponent is directly evaluated from the early-time 
relaxation of magnetization, and compared with the results in sections 
III and V. The above descriptions are summarized in section VII.

\section{Formulation}
In the present article we investigate the two- and three-dimensional 
Ising models with the nearest-neighbor interaction on a square and 
simple cubic lattices, respectively, 
\begin{equation}
{\cal H}=-J \sum_{\langle ij \rangle \in {\rm n.n.}} S_{i}S_{j},\ S_{i} = \pm 1, 
\end{equation}
with the Swendsen-Wang (SW) algorithm~\cite{SW}. 
We have already found~\cite{Nonomura14,Nonomura16} 
that the stretched-exponential behavior is observed
at the critical temperature $T_{\rm c}$ both in the 
decaying process from the perfectly-ordered state and 
the ordering process from the perfectly-disordered state, 
and that initial-time deviation in the decaying process 
is much larger than that in the ordering process. 
In the present article, we therefore concentrate on the 
ordering process from the perfectly-disordered state. 

Early-time behavior of the absolute value of 
the magnetization at $T_{\rm c}$ is given by 
\begin{equation}
\label{se-dis}
\langle |m(t,L)| \rangle \sim L^{-d/2} \exp \left( + c_{m} t ^{\sigma} \right)\ 
(0<\sigma <1),
\end{equation}
with the spatial dimension $d$, a relaxation constant $c_{m}$ possibly 
depending on observed quantities and the exponent $\sigma$ 
independent of quantities. Here the explicit size dependence 
originates from the normalized random-walk growth of clusters. 
Similar behavior is also observed in the squared magnetization, 
\begin{equation}
\label{se-m2}
\langle m^{2}(t,L) \rangle \sim L^{-d} \exp \left( + c_{m^{2}} t ^{\sigma} \right),
\end{equation}
and the $(2,1)$-Binder ratio is therefore scaled as 
\begin{eqnarray}
\label{Bin}
B_{{2,1}}(t,L) &\equiv& \langle m^{2}(t,L) \rangle / \langle |m(t,L)| \rangle^{{2}} \nonumber\\
                      &\sim&  \exp \left[ + (c_{m^{2}} - c_{m}^{{2}}) t ^{\sigma} \right].
\end{eqnarray}

Similarly, early-time behavior of the correlation length at $T_{\rm c}$ 
is expressed as 
\begin{equation}
\label{xi-eq}
\xi(t,L) \sim \exp \left( +c_{\xi} t^{\sigma} \right),
\end{equation}
while this quantity is scaled with $\sim L^{\nu/\nu}=L$ in equilibrium. 
Then, taking $\rho \equiv 1/c_{\xi}$ and the size-independent 
quantity $\xi(t,L)/L$ is scaled as~\cite{Tomita17} 
\begin{eqnarray}
\label{xi-sc}
\xi(t,L)/L
&\sim& \exp \left( +\rho^{-1}t^{\sigma} -\ln L \right) \nonumber\\*
&\sim& \exp \left[ +\rho^{-1}\left( t^{\sigma}-\ln L^{\rho} \right) \right].
\end{eqnarray}
This relation indicates that $\xi(t,L)/L$ is scaled well for various 
system sizes with the scaling quantity $t^{\sigma} - \ln L^{\rho}$ 
at least for the both ends of this quantity, and actually in the 
whole parameter region as shown numerically~\cite{Tomita17}. 
Although this functional form of the scaling quantity resembles 
that of the nonequilibrium-to-equilibrium scaling observed in 
size-dependent quantities~\cite{Nonomura14,Nonomura16}, 
origin of the exponent $\rho$ is not the same, and such a 
scaling in $\xi(t,L)/L$ may also hold in $B_{2,1}(t,L)$. 
Note that $\rho$ may depend on quantities because 
the coefficient $c$ in Eq.~(\ref{xi-eq}) may depend 
on them, while the relaxation exponent $\sigma$ is 
more fundamental and independent of quantities.

\section{Numerical results in the two-dimensional Ising model}
In the present section, we explain the procedure to evaluate 
critical phenomena (especially the relaxation exponent $\sigma$) 
on the basis of the Binder ratio in the two-dimensional Ising model 
on a square lattice.

In our previous studies, we evaluated the critical temperature $T_{{\rm c}}$ 
and critical exponents with the nonequilibrium-to-equilibrium scaling, where 
the initial-time stretched-exponential scaling form (such as Eq.~(\ref{se-dis})) 
and the finite-size scaling form in equilibrium 
(such as $m_{{\rm c}}(L)\sim L^{{-\beta/\nu}}$) are coupled. 
Since such a scheme is essentially a four-parameter fitting 
($T_{\rm c}$, $\beta/\nu$, $\sigma$ and $c_{m}$), 
precision of the parameters was rather limited.

\begin{figure}
\includegraphics[width=88mm]{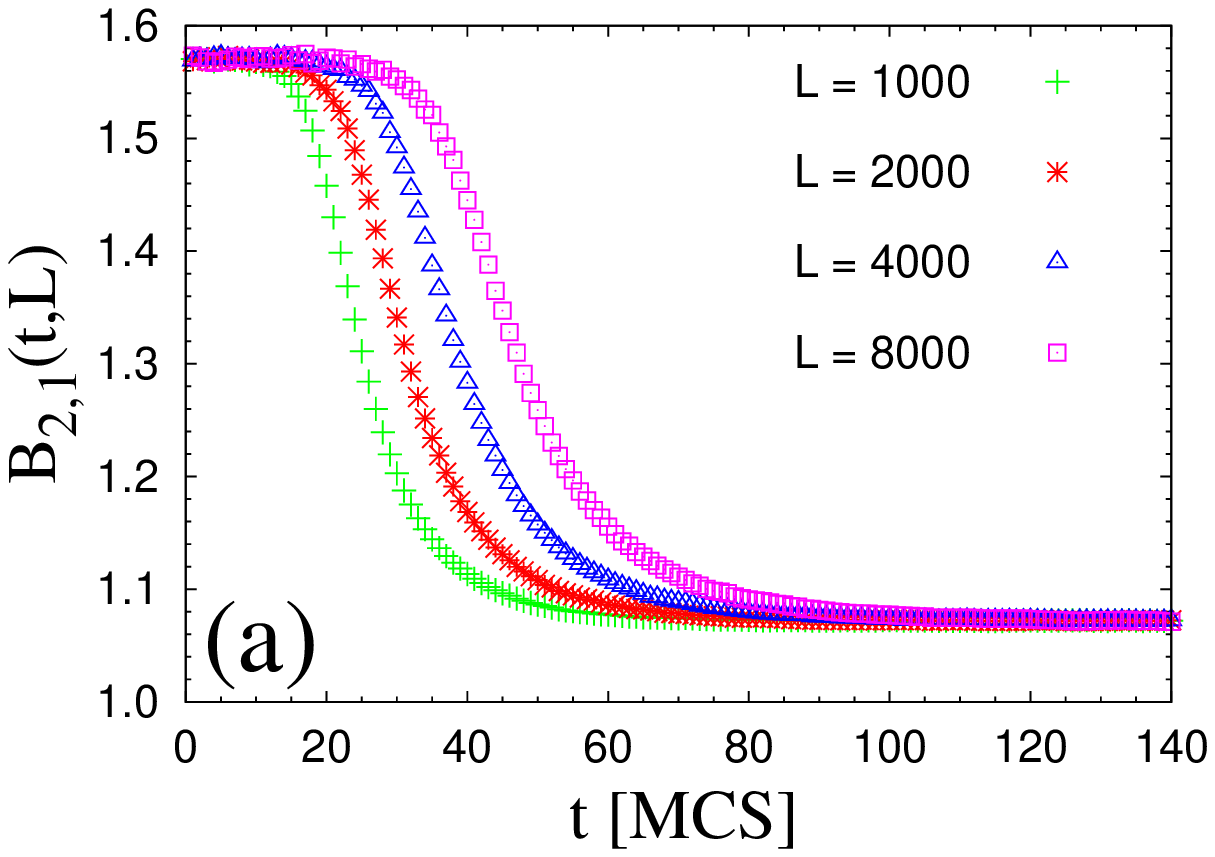}
\includegraphics[width=88mm]{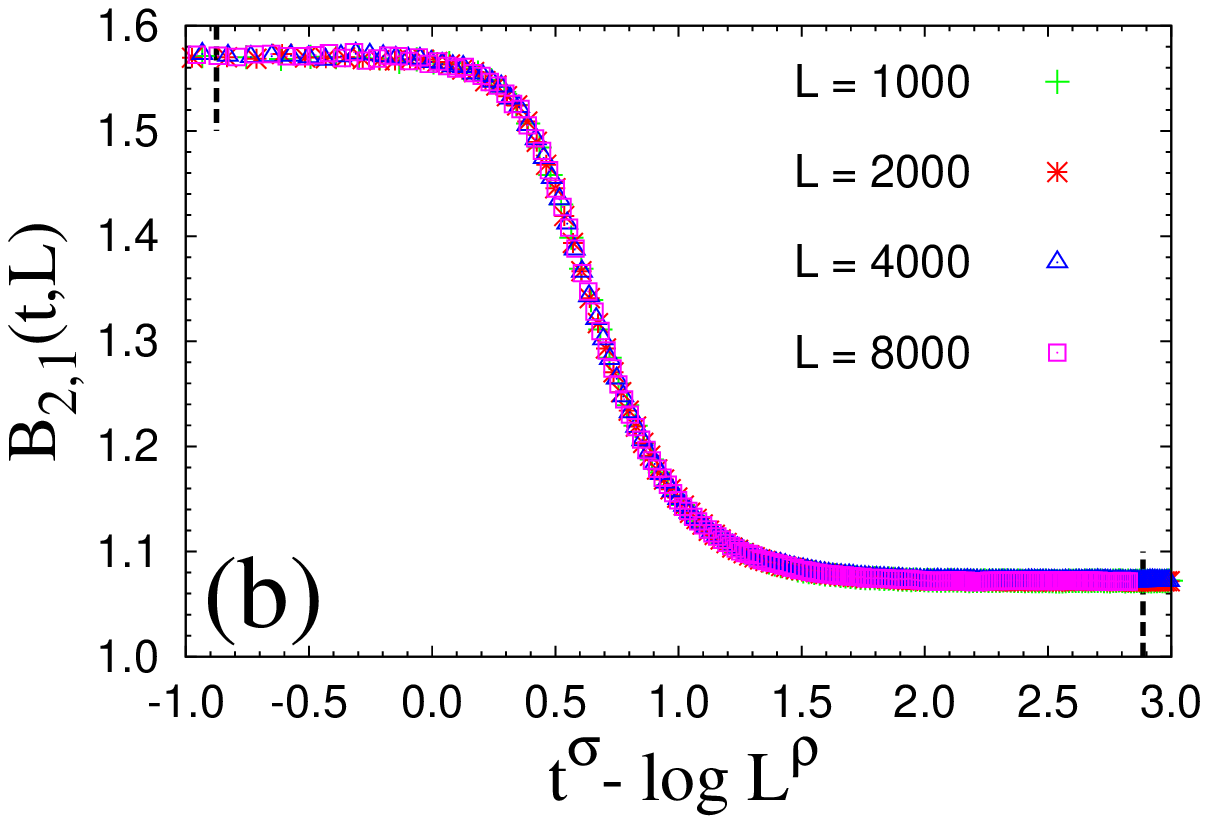}
\caption{(Color online) (a) Simulation-time dependence 
of the $(2,1)$-Binder ratio at the exact $T_{\rm c}$ for 
$L=1000$ (crosses), $2000$ (stars), $4000$ (triangles) 
and $8000$ (squares). No size dependence is observed 
at both ends of time evolution. (b) Semi-empirical scaling plot 
of this quantity with $\sigma=0.3277(1)$ and $\rho=0.3168(3)$ 
based on the data inside of the dashed lines.
}
\label{br-2D}
\end{figure}
Here we propose a new scheme to evaluate such parameters 
more precisely. Since this scheme is based on a semi-empirical 
scaling form following Eq.~(\ref{xi-sc}), we use the exact value, 
$T_{\rm c} = 2/\log ( 1 + \sqrt{2} ) {\rm [} J/k_{\rm B} {\rm ]}
= 2.2691853\ldots {\rm [} J/k_{\rm B} {\rm ]}$, 
in order to avoid extra efforts for justification of the scaling form. 
Evaluation process of $T_{\rm c}$ will be given in Section V 
for the three-dimensional case.

First, the $(2,1)$-Binder ratio at $T_{\rm c}$ is plotted versus simulation time 
in Fig.\ \ref{br-2D}(a) for $L=1000$ ($6.4 \times 10^{5}$ random number 
sequences (RNS) are averaged), $2000$ ($3.2 \times 10^{5}$ RNS), 
$4000$ ($1.6 \times 10^{5}$ RNS) and $8000$ ($0.8 \times 10^{5}$ RNS). 
As expected, this quantity becomes size independent as the system 
approaches equilibrium. Moreover, it also seems size independent at 
the onset of simulations (origin of this behavior will be explained in the 
next section), and even its initial slope of time evolution looks vanishing, 
in spite of the expected stretched-exponential time dependence (\ref{Bin}). 
Actually, this cancellation, $c_{m^{2}}-c_{m}^{2}=0$, is specific 
to the Ising models. When classical vector spin models 
are simulated~\cite{Nonomura15,Nonomura16} with 
the embedded-Ising-spin algorithm~\cite{Wolff}, such 
cancellation is not observed~\cite{Nono_progress}.

Although such size independence is suitable for evaluation of $T_{\rm c}$, 
the exponent $\sigma$ does not appear explicitly in the expression 
of $B_{2,1}(t,L)$ anymore. Nevertheless, this quantity consists of 
$\langle |m(t,L)| \rangle$ and $\langle m^{2}(t,L) \rangle$, and 
they show the stretched-exponential relaxation as given in 
Eqs.~(\ref{se-dis}) and (\ref{se-m2}). 
Then, the information of $\sigma$ would remain in $B_{2,1}(t,L)$ 
as a higher-order correction. In Fig.~\ref{br-2D}(b), this quantity 
is plotted versus a rescaled time $t^{\sigma}-\ln L^{\rho}$ with 
$\sigma=0.3277(1)$ and $\rho=0.3168(3)$. These exponents 
are evaluated so as to minimize the mutual residue of the data. 
The ranges of the data used for the fitting (represented by dashed 
lines in Fig.\ \ref{br-2D}(b)) are also determined by minimizing the 
residue. Since the residue decreases monotonically as the upper 
range is increased, it is fixed so as to include all the data for 
$L=8000$. Note that the estimate of $\sigma$ is not inconsistent 
with the expected value, $\sigma=1/3$~\cite{Nonomura14}. 

\section{Distribution function \hspace{2.0cm} 
of magnetization in the two-dimensional Ising model}

In the present section, we analyze the distribution function 
of magnetization $P(|m|)$ in the two-dimensional Ising 
model at $T_{\rm c}$ and clarify physical background 
of the stretched-exponential critical relaxation and 
simulation-time dependence of the $(2,1)$-Binder ratio. 

\begin{figure}
\includegraphics[width=88mm]{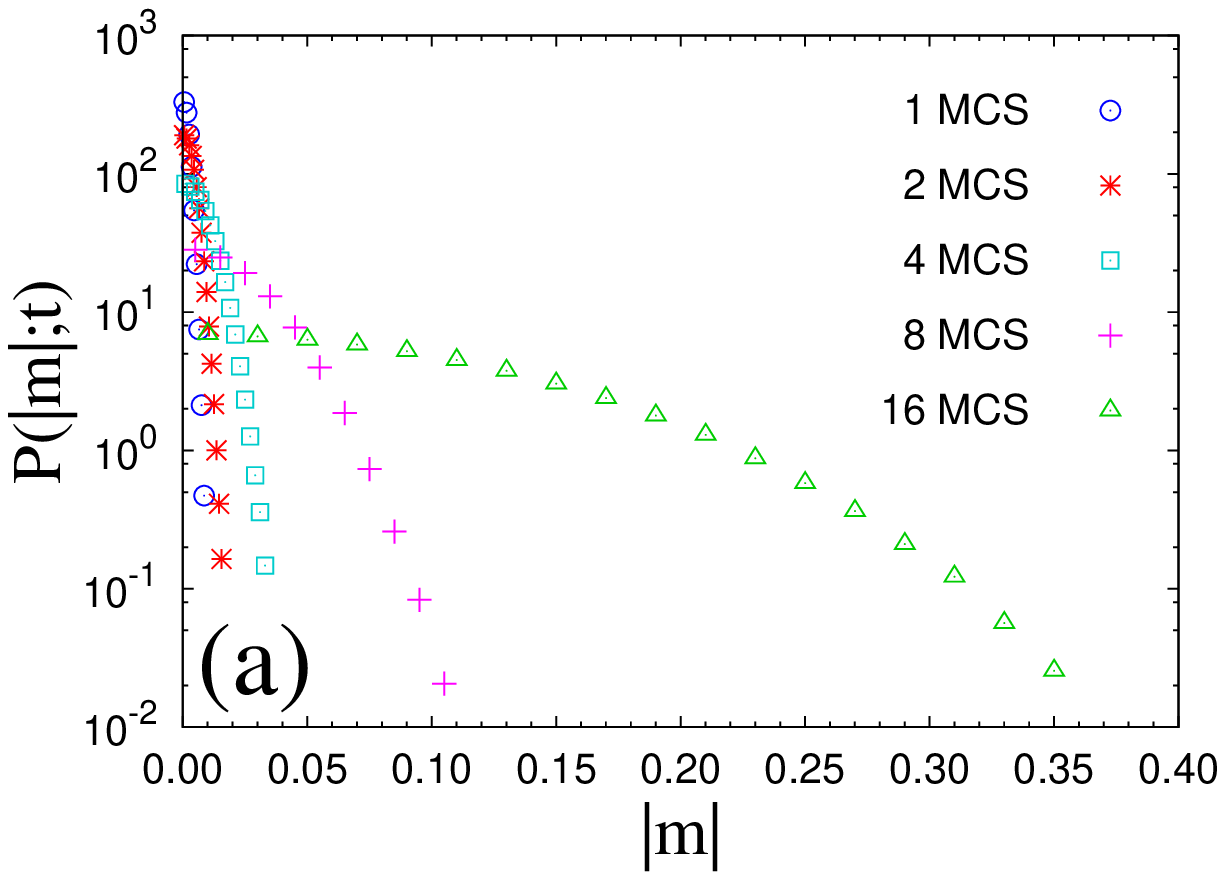}
\includegraphics[width=88mm]{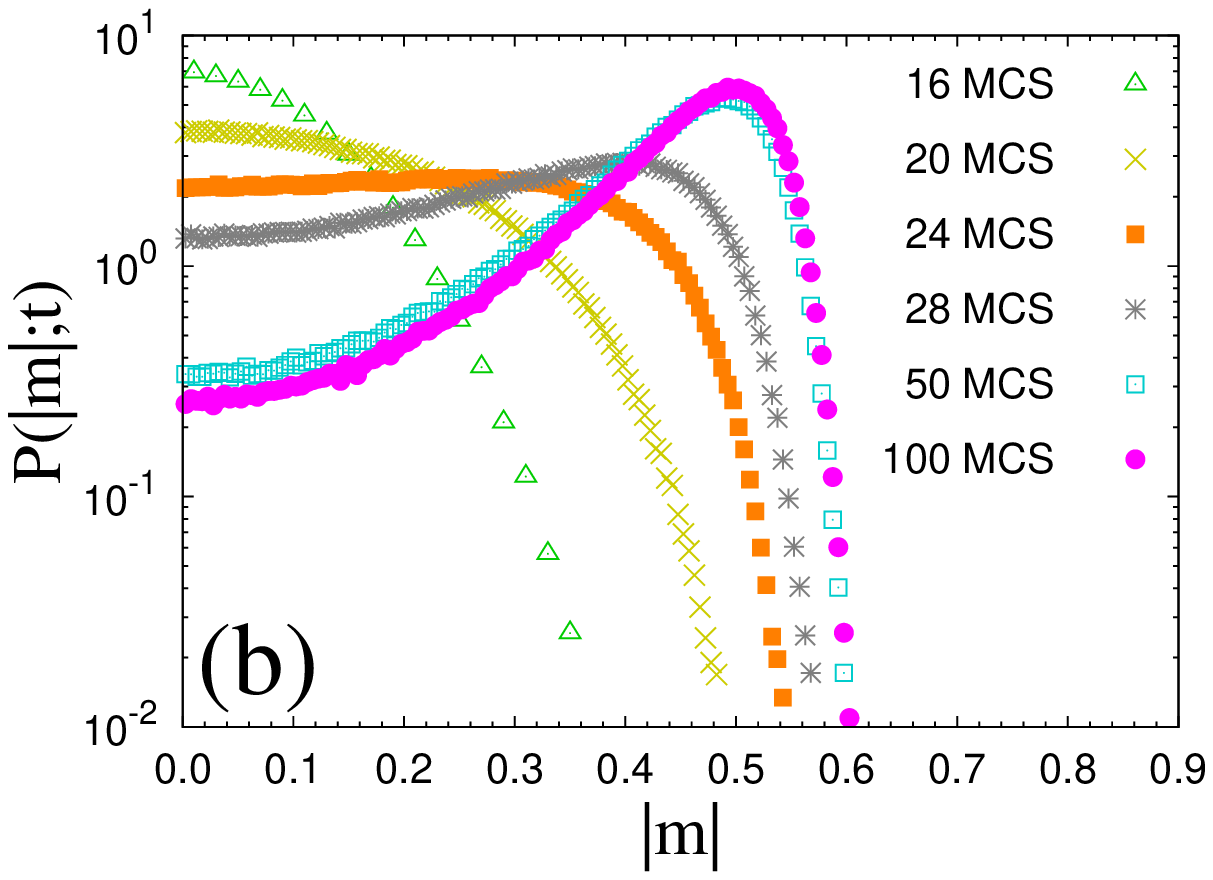}
\caption{(Color online) Distribution function of magnetization 
at $T_{\rm c}$ for $L=1000$ at various Monte Caro steps. 
(a) Data in the earlier stage at 1MCS (open circles), 2MCS (stars), 
4MCS (open squares), 8MCS (crosses) and 16MCS (open triangles). 
(b) Data in the later stage at 16MCS (open triangles), 20MCS (X-marks), 
24MCS (full squares), 28MCS (stars), 50MCS (open squares) and 
100MCS (full circles).
}
\label{dis-2D}
\end{figure}
In Fig.~\ref{dis-2D}, the distribution function of magnetization $P(|m|)$ 
is plotted versus the absolute value of magnetization $|m|$ at various 
Monte Carlo steps (MCS) for $L=1000$. Data in the earlier stage 
(at 1, 2, 4, 8 and 16MCS) and in the later stage (at 16, 20, 24, 28, 50 and 
100MCS) are shown in Figs.~\ref{dis-2D}(a) and \ref{dis-2D}(b), respectively. 
This function is obtained from $6.4 \times 10^{5}$ samples with different RNS 
and divided into $10,000$ meshes for $0 \leq |m| \leq 1$. It is normalized as 
$\int_{0}^{1}P(|m|){\rm d}|m|=1$, and the data points in these figures are 
truncated for clear visualization. In the earlier stage, $P(|m|)$ has a dome-like 
shape with the peak at $|m|=0$, and in the later stage it gradually approaches 
the equilibrium distribution with the peak around the critical magnetization 
$m_{\rm c}(L)$. 

\begin{figure}
\includegraphics[width=88mm]{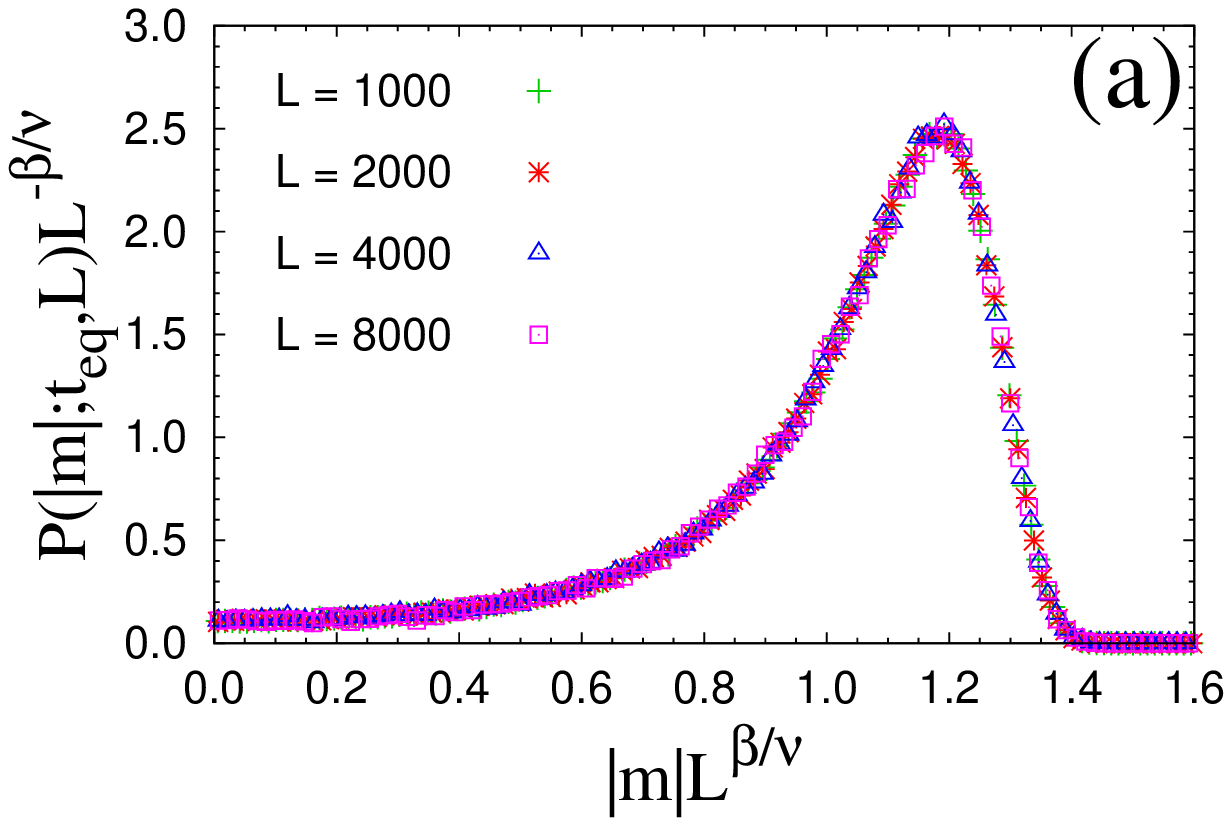}
\includegraphics[width=88mm]{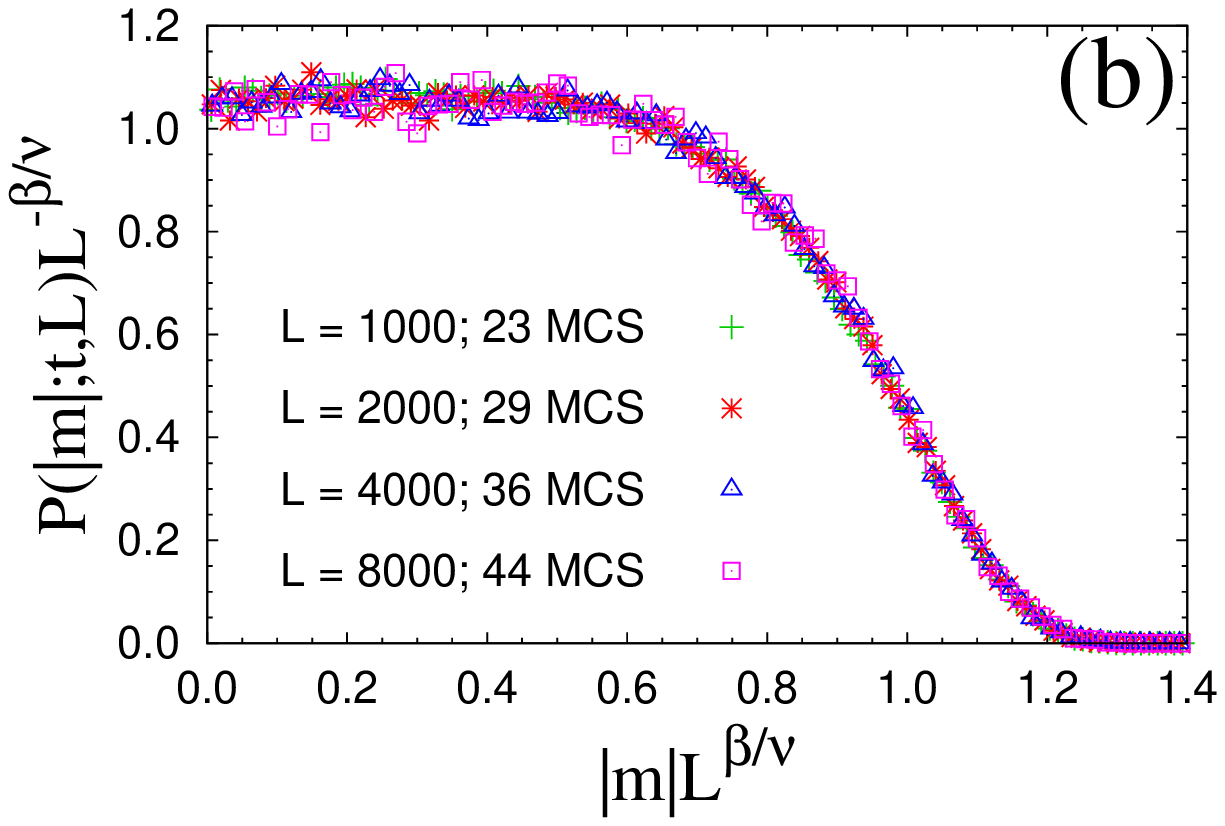}
\caption{(Color online) Scaling behavior of the distribution function 
of magnetization at $T_{\rm c}$, where $P(|m|;t,L)L^{-\beta/\nu}$ 
is plotted versus $|m|L^{\beta/\nu}$ with $\beta/\nu=1/8$ for $L=1000$ 
(crosses), $2000$ (stars), $4000$ (triangles) and $8000$ (squares) 
(a) in equilibrium and (b) at the moment with the plateau-like 
distribution, namely at 23MCS for $L=1000$, 29MCS for 
$L=2000$, 36MCS for $L=4000$ and 44MCS for $L=8000$.
}
\label{dis-sc}
\end{figure}
Then, $P(|m|)$ is scaled with system sizes. In Fig.~\ref{dis-sc}(a), it is scaled in 
equilibrium (actually at $t_{\rm eq}=200$MCS) for $L=1000$, $2000$, $4000$ and 
$8000$. Here $P(|m|) L^{-\beta/\nu}$ is scaled with $|m| L^{\beta/\nu}$~\cite{Tomita99} 
with the exact critical exponent $\beta/\nu=1/8$, and this behavior is consistent 
with the fact that the peak of $P(|m|)$ in equilibrium corresponds to the critical 
magnetization, $m_{\rm c}(L) \sim L^{-\beta/\nu}$. This behavior also holds 
during the relaxation process. In Fig.~\ref{dis-sc}(b), $P(|m|)$ with a plateau-like 
distribution is scaled with the same formula for $L=1000$ (at 23MCS), $2000$ (at 
29MCS), $4000$ (at 36MCS) and $8000$ (at 44MCS). Although similar scaling 
behavior is also expected to be observed at other moments, it is generally 
difficult to take corresponding configurations for different system sizes. 

\begin{figure}
\includegraphics[width=88mm]{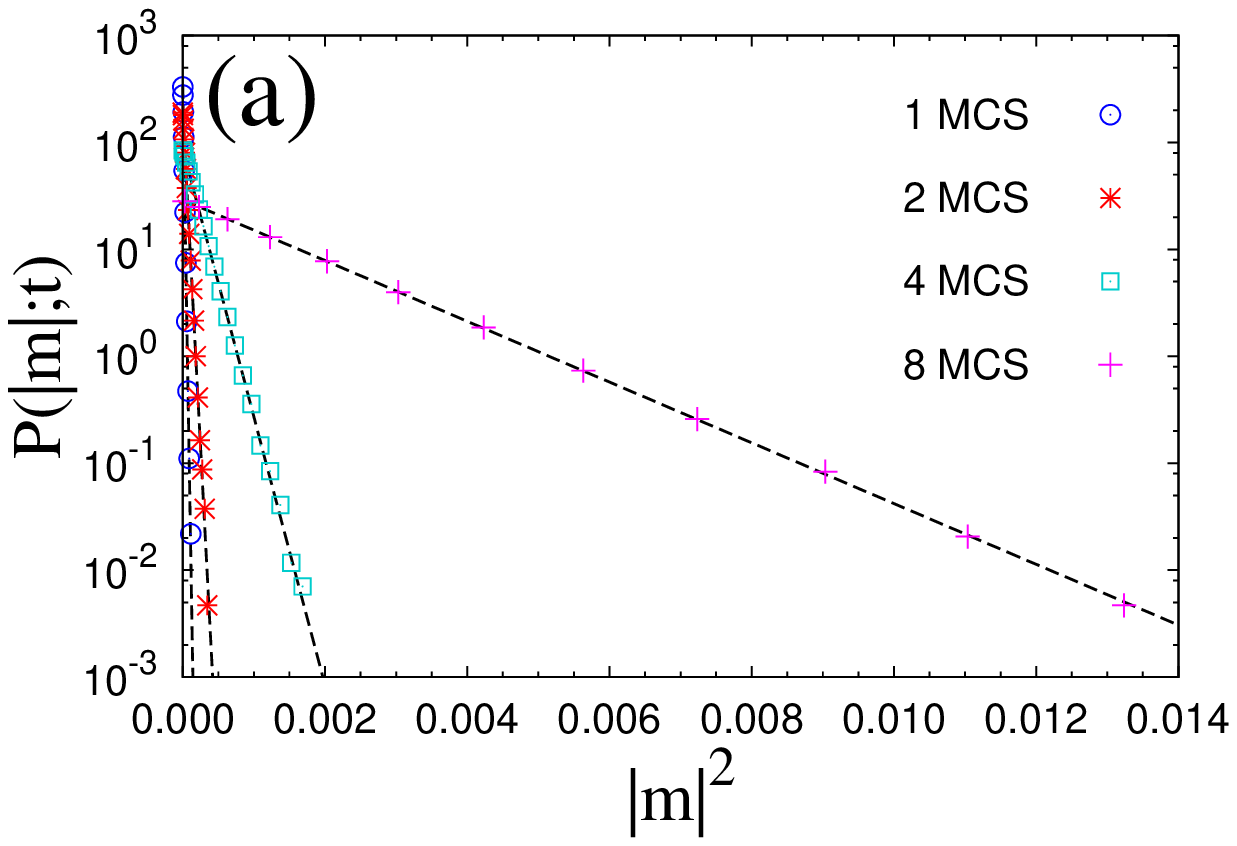}
\includegraphics[width=88mm]{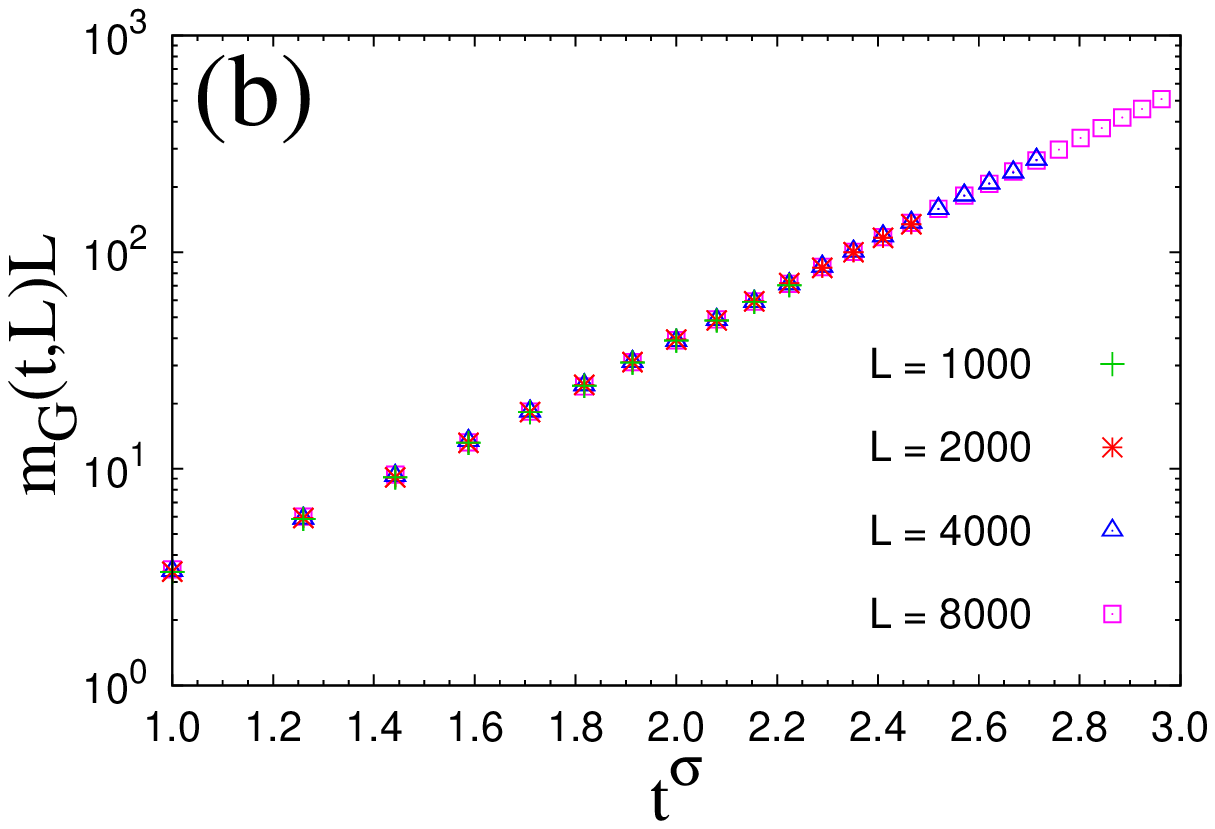}
\caption{(Color online) (a) Distribution function of magnetization 
versus square of magnetization for $L=1000$ at $1$MCS (circles), 
$2$MCS (stars), $4$MCS (squares) and $8$MCS (crosses), which 
gives the Gaussian distribution function (\ref{Gauss}). The dashed 
lines stand for the fitting curves based on Eq.~(\ref{Gauss}).
(b) Scaling behavior of the width of the Gaussian distribution function. 
Here $m_{\rm G}(t,L)L$ plotted versus $t^{\sigma}$ with $\sigma=1/3$ 
in a semi-log scale shows a linear behavior for $L=1000$ (crosses), 
$2000$ (stars), $4000$ (triangles) and $8000$ (squares), 
which results in the scaling form (\ref{se-Gauss}). 
}
\label{Gaussian}
\end{figure}
Next, we consider the scaling behavior at the onset of relaxation. 
Although similar scaling analyses become further difficult, a 
square-$|m|$ plot of $P(|m|;t)$ (Fig.~\ref{Gaussian}(a) for 
$L=1000$) results in the Gaussian distribution function, 
\begin{equation}
\label{Gauss}
P(|m|;t,L) \sim \exp \left[ -(|m|/m_{\rm G}(t,L))^{2} \right].
\end{equation}
That is, the early-time relaxation behavior is described by the Gaussian 
distribution, and the size-independent initial value of $B_{2,1}(t=0)$ is 
derived from the Gaussian integrals $\langle \cdots \rangle_{\rm G}$, 
\begin{equation}
B_{2,1}(t=0) \equiv \frac{\langle m^{2}(t=0) \rangle_{\rm G}}
                                       {\langle |m(t=0)| \rangle_{\rm G}^{{2}}}
=\frac{m_{\rm G}^{2}/2}{\left( m_{\rm G} / \sqrt{\pi} \right)^{2}}=\frac{\pi}{2},
\end{equation}
and numerical data of $P(|m|;t,L)$ for various system sizes and 
Monte Carlo steps are fitted with Eq.~(\ref{Gauss}) to evaluate 
$m_{\rm G}(t,L)$ as shown in Fig.~\ref{Gaussian}(b). 
Here $m_{\rm G}(t,L)L$ is plotted in a semi-log scale versus 
$t^{\sigma}$ with $\sigma=1/3$, and such linear behavior 
suggests the following scaling form, 
\begin{equation}
\label{se-Gauss}
m_{\rm G}(t,L) \sim L^{-1} \exp \left( Ct^{\sigma} \right), \ \sigma=1/3.
\end{equation}

We can obtain various insights from the above results. First, we find 
that the stretched-exponential relaxation of magnetization (\ref{se-dis}) 
originates from the simulation-time dependence of the width of the 
Gaussian distribution of magnetization (\ref{se-Gauss}), which 
suggests that the scaling behavior (\ref{se-dis}) is fundamental 
in the SW algorithm. Second, the data in Fig.~\ref{Gaussian}(b) 
reveal that the width of the Gaussian distribution function shrinks 
with $\sim L^{-1}$, which is much faster than the scaling of the 
peak value of $P(|m|;t,L)$ in equilibrium (Fig.~\ref{dis-sc}(a)) 
or the width of it at the moment with a plateau-like distribution 
(Fig.~\ref{dis-sc}(b)) with $\sim L^{-\beta/\nu}$. The power $-1$ 
in Eq.~(\ref{se-Gauss}) would be identified with $-d/2$, which 
characterizes the random-walk growth of magnetization as seen 
in Eq.~(\ref{se-dis}). 
Third, the data for each system size in Fig.~\ref{Gaussian}(b) 
are plotted up to the limit where the distribution function is 
described well with the Gaussian formula, and this limit 
coincides with that of early-time independence of 
$B_{2,1}(t,L)$ displayed in Fig.~\ref{br-2D}(a). 

\begin{figure}
\includegraphics[width=88mm]{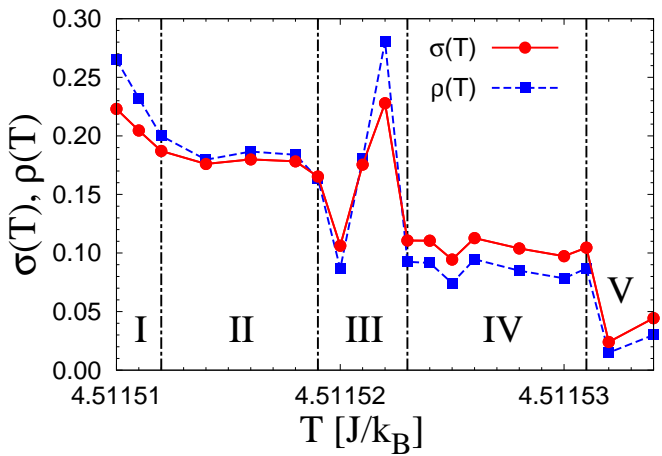}
\caption{(Color online) Temperature dependence of the exponents 
$\sigma(T)$ (circles) and $\rho(T)$ (squares). Evaluation process of 
these exponents is shown in  Fig.~\ref{3D-fit} for some temperatures. 
Five temperature regions (I to V) are explained in the text. 
}
\label{3Dexp}
\end{figure}
\begin{figure*}
\includegraphics[width=88mm]{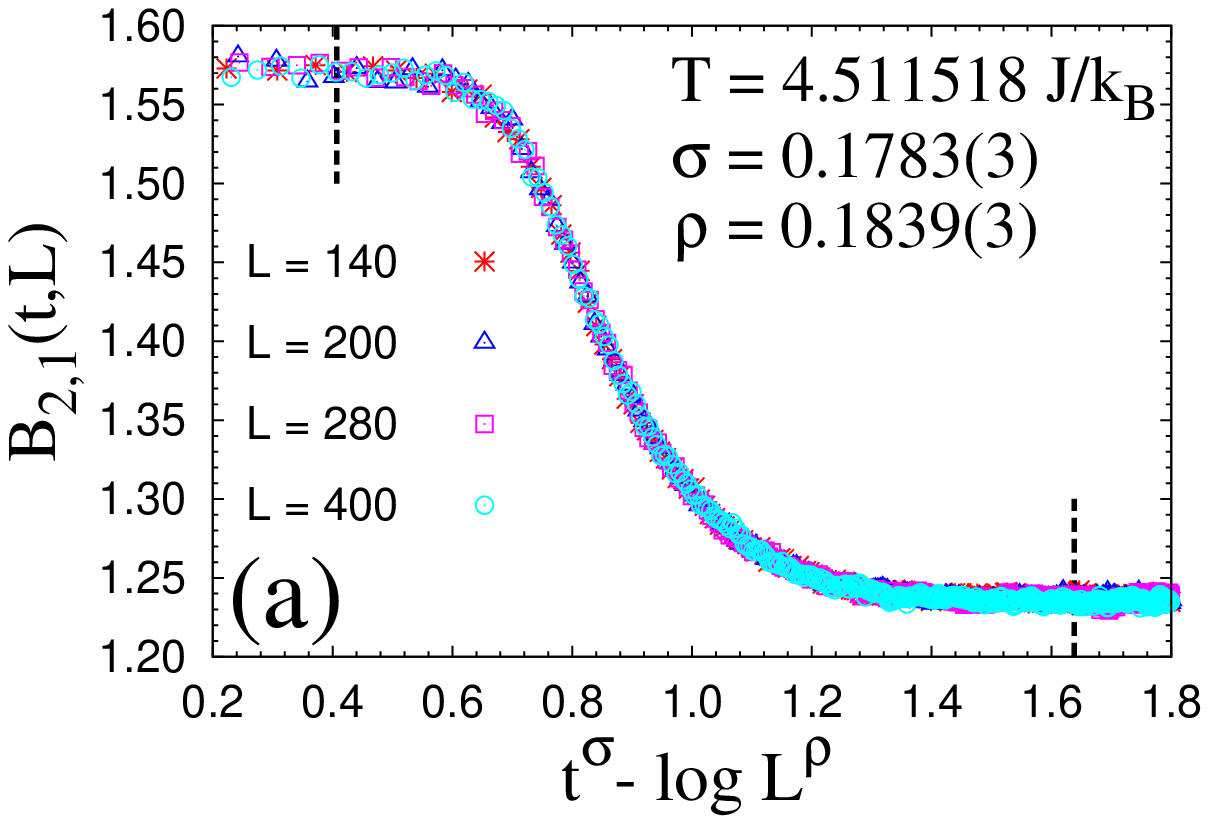}
\includegraphics[width=88mm]{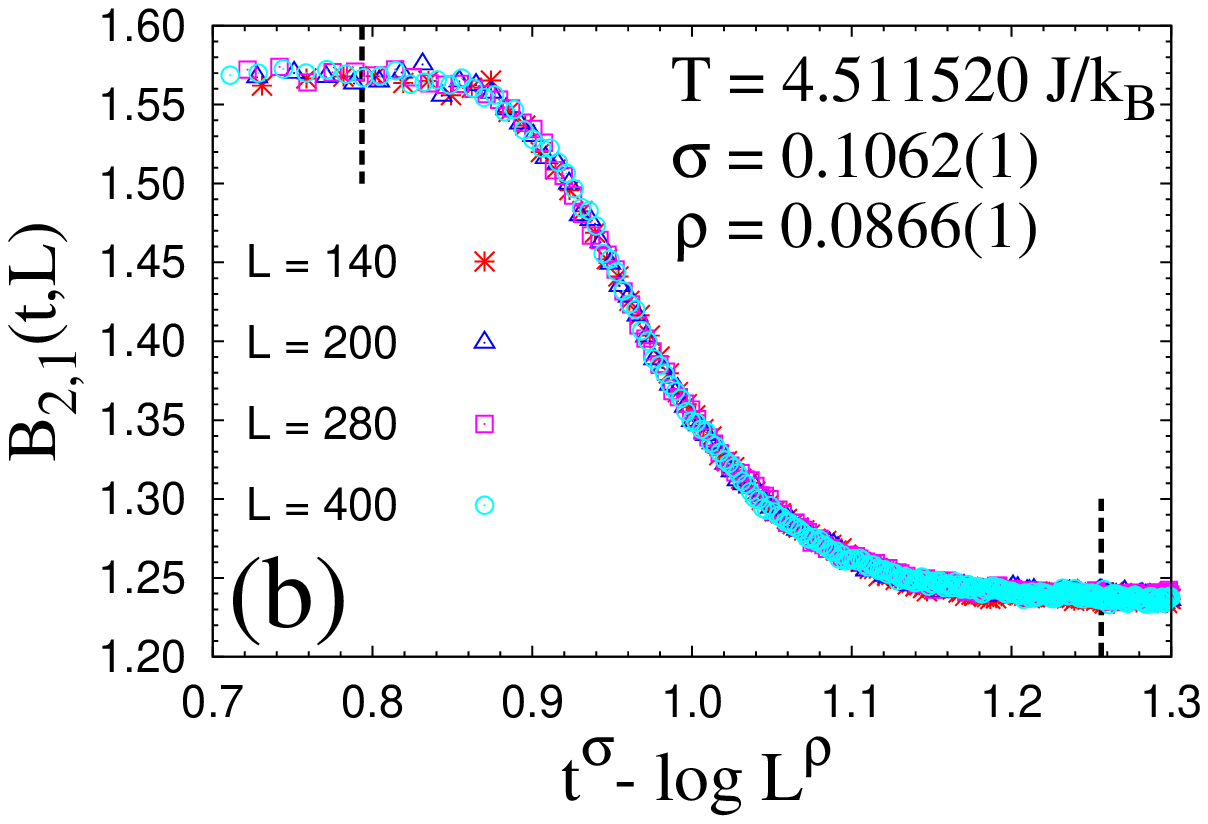}
\includegraphics[width=88mm]{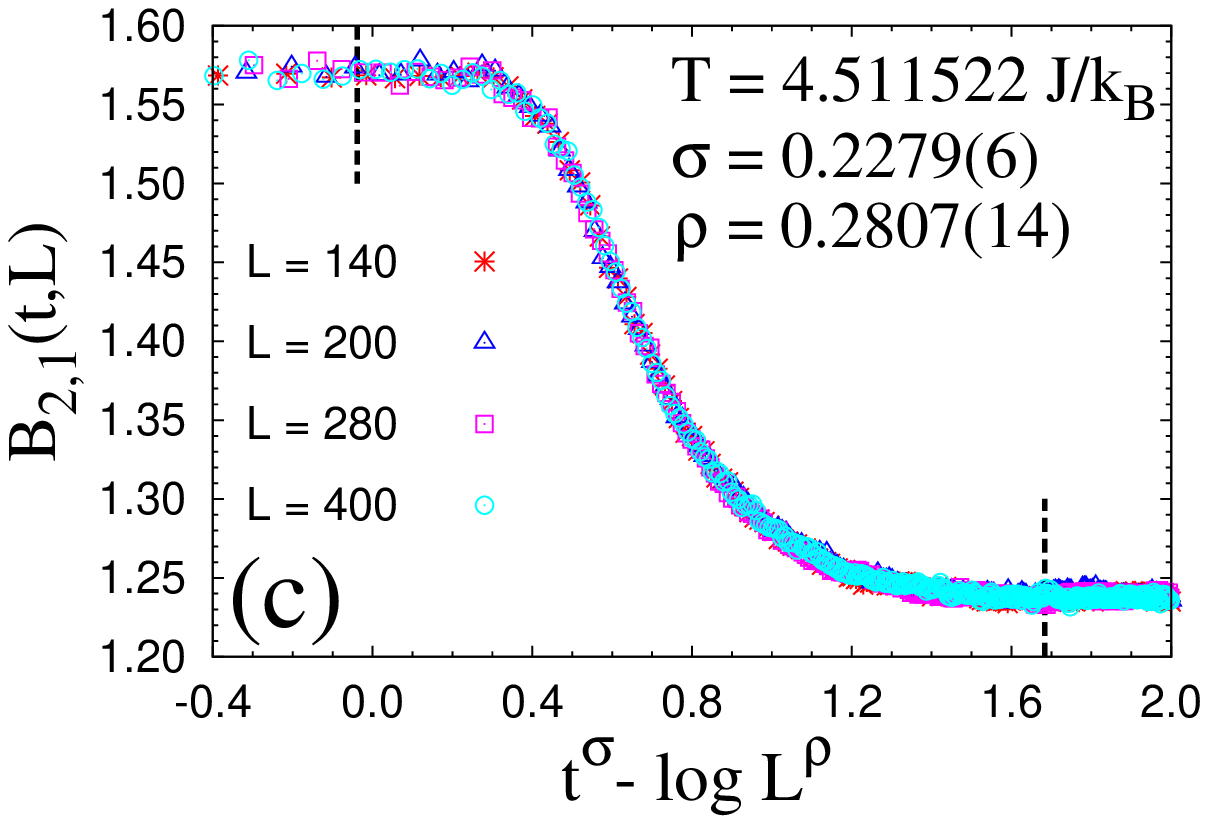}
\includegraphics[width=88mm]{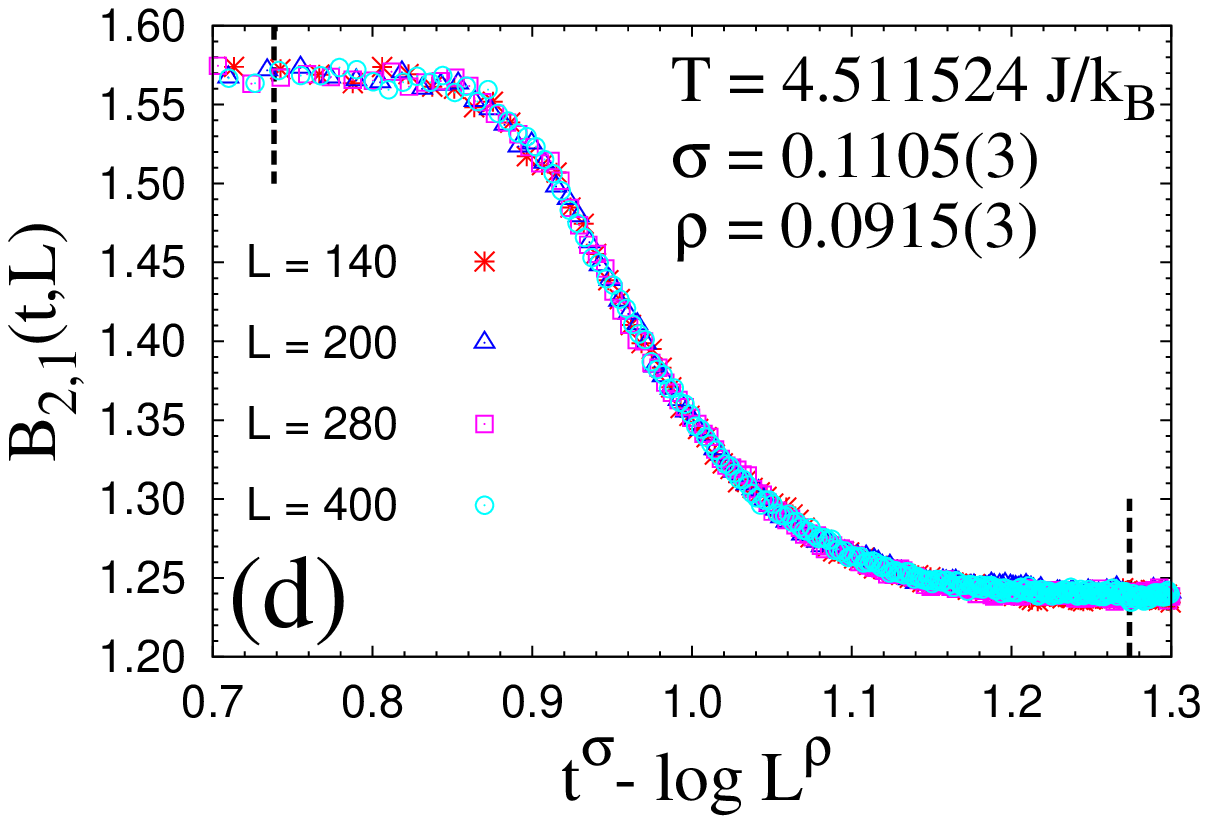}
\caption{(Color online) Semi-empirical scaling plot 
of the $(2,1)$-Binder ratio based on the data 
inside of the dashed lines for $L=140$ (stars), 
$200$ (triangles), $280$ (squares) and $400$ (circles) at 
(a) $T=4.511518 J/k_{\rm B}$, (b) $4.511520 J/k_{\rm B}$, 
(c) $4.511522 J/k_{\rm B}$ and (d) $4.511524 J/k_{\rm B}$. 
The estimates of $\sigma(T)$ and $\rho(T)$ at each 
temperature are exhibited in each figure.
}
\label{3D-fit}
\end{figure*}
\section{Numerical results in the three-dimensional Ising model}
In the present section, we evaluate critical phenomena of the 
three-dimensional Ising model on a simple cubic lattice by the 
scheme explained in Section III. Here we vary the temperature around 
the critical region, and show that the critical temperature $T_{\rm c}$ 
can be evaluated very accurately from the temperature dependence 
of the relaxation exponent $\sigma(T)$, which is of interest in itself. 

Temperature dependence of the relaxation exponents 
$\sigma(T)$ and $\rho(T)$ evaluated from the scaling plot 
similarly to the one in Fig.~\ref{br-2D}(b) is displayed 
In Fig.~\ref{3Dexp}. Some examples of the fitting of 
$B_{2,1}(t,L)$ versus $t^{\sigma}-\ln L^{\rho}$ for 
$L=140$, $200$, $280$ and $400$ ($4.0 \times 10^{4}$ 
RNS are averaged in all the sizes and temperatures) are 
given in Figs.\ \ref{3D-fit}(a)-(d) at $T=4.511518 J/k_{\rm B}$, 
$4.511520 J/k_{\rm B}$, $4.511522 J/k_{\rm B}$ 
and $4.511524 J/k_{\rm B}$, respectively. 
The range of the temperature shown in Fig.~\ref{3Dexp} 
is consistent with an estimate of the critical temperature 
based on the conventional NER analysis using 
the $6648 \times 6648 \times 6656$ cluster, 
$J/k_{\rm B}T_{\rm c}=0.2216547(5)$ or 
$T_{\rm c}=4.511522(10) J/k_{\rm B}$~\cite{Ito05} . 

The temperature region shown in Fig.~\ref{3Dexp} is divided into 
five subregions. For $T < 0.4511512 J/k_{\rm B}$ (region I), the 
relaxation exponent $\sigma(T)$ increases as $T$ decreases. 
For $4.511512 J/k_{\rm B} \leq T \leq 4.511519 J/k_{\rm B}$ (region II), 
the exponent $\sigma(T)$ almost takes a constant value $\sigma = 0.178(2)$. 
For $4.511519 J/k_{\rm B}<T<4.511523 J/k_{\rm B}$ (region III), 
$\sigma(T)$ exhibits a sharp temperature dependence. 
For $4.511523 J/k_{\rm B} \leq T \leq 4.511531 J/k_{\rm B}$ 
(region IV), $\sigma(T)$ almost takes a constant value again, 
$\sigma = 0.107(4)$. For $T > 4.511531 J/k_{\rm B}$ (region V), 
$\sigma(T)$ is much smaller.

Physical interpretation of these results is as follows: 
The regions between II and IV may correspond to 
the critical one, and the critical temperature can be 
estimated as $T_{\rm c}=4.511521(2) J/k_{\rm B}$. 
Rapidly-changing value of $\sigma(T)$ indicates 
large fluctuations around $T_{\rm c}$, and the above 
estimate of $T_{\rm c}$ is identified with the region III. 
The estimate $\sigma = 0.178(2)$ in the region II 
(just below $T_{\rm c}$) would stand for the true 
relaxation exponent. On the other hand, the estimate 
$\sigma = 0.107(4)$ in the region IV (just above 
$T_{\rm c}$) might be a fictitious one in finite systems, 
which signals the stretched-exponential relaxation still 
holds in this region. Large discrepancies of $\sigma(T)$ 
in the regions I and V simply tell that these regions are 
off-critical and the stretched-exponential relaxation does 
not hold anymore. In the region I, the system is in the 
ordered phase and the magnetization grows exponentially, 
and $\sigma(T)$ is expected to approach unity as $T$ 
decreases. In the region V, the system is in the paramagnetic 
phase and no magnetic long-range order is stable.

Note that the exponents $\sigma(T)$ and $\rho(T)$ are 
not so different around $T_{\rm c}$, and they almost 
coincide with each other in the region II. Actually, their 
relation in sizes is reversed in the regions II and IV, and 
they seem to converge at $\sigma \approx \rho \approx 1/6$ 
(e.g.~at $T=4.511519 J/k_{\rm B}$). A similar relation also 
exists in two dimensions, where such convergence of the 
two exponents occurs at $\sigma \approx \rho \approx 1/3$. 
\section{Comparison of the relaxation exponent $\sigma$ 
directly obtained from stretched-exponential relaxation}
In the previous section, a large temperature dependence of the ``relaxation 
exponent" $\sigma(T)$ is observed in the region III in Fig.~\ref{3Dexp} in the 
three-dimensional Ising model. Then, it is interesting to investigate whether 
this large fluctuation is actually observed in physical quantities or a fictitious 
behavior specific to the analysis based on the Binder ratio. For this purpose, 
direct observation of the exponent $\sigma$ defined in Eq.~(\ref{se-dis}) is 
straightforward. 

\subsection{Two-dimensional Ising model at $T_{\rm c}$}
\begin{figure}
\includegraphics[width=88mm]{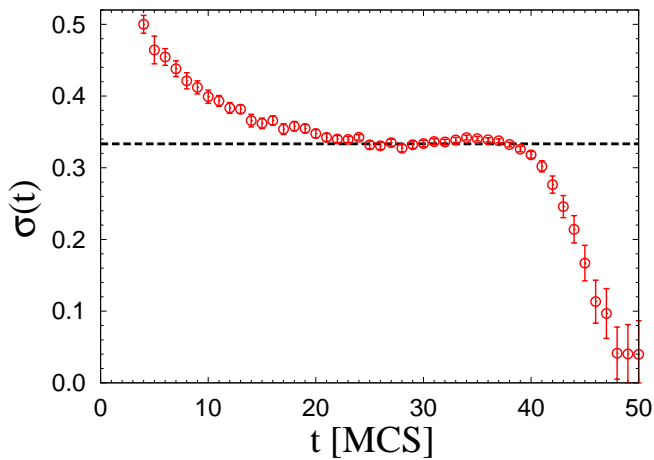}
\caption{(Color online) Relaxation exponent $\sigma(t)$ evaluated from the 
least-squares fitting of the initial $t$-MCS data of $\langle |m(t)| \rangle$ 
for $L=8000$ at $T_{\rm c}$ in the two-dimensional Ising model based 
on Eq.~(\ref{se-dis}) versus $t$ used for the fitting. The dashed line 
corresponds to $\sigma=1/3$ as a guide for eyes.
}
\label{sig2D}
\end{figure}
Consequence of the above observation in the two-dimensional Ising 
model at $T_{\rm c}$ for $L=8000$ is displayed in Fig.~\ref{sig2D}. 
Here relaxation data of $\langle |m(t)| \rangle$ from $1$ to $t$ MCS 
are fitted with Eq.~(\ref{se-dis}), and the estimate of $\sigma(t)$ 
is plotted versus $t$ used for the fitting. For larger $t$, the 
stretched-exponential relaxation has already been saturated 
(it is signaled by the drop of $B_{2,1}(t,L)$ from the Gaussian 
value $\pi/2$ in Fig.~\ref{br-2D}(a)), and the fitting based on 
Eq.~(\ref{se-dis}) becomes poor (indicated by large error bars) 
and the estimate of $\sigma(t)$ deviates rapidly as $t$ increases. 

For $L=8000$, $\sigma(t)$ monotonically decreases as $t$ increases 
up to $t=26$, which is the upper limit of the Gaussian behavior of 
the distribution function of $|m|$ as shown in Fig.~\ref{Gaussian}(b). 
The residue for fitting of the data between $1$ to $26$ MCS takes 
minimum and we have $\sigma=0.330(5)$, which is both consistent 
with the estimate from $B_{2,1}(t,L)$, $\sigma=0.3277(1)$ and 
the expected value $\sigma=1/3$. For $27 \lesssim t \lesssim 40$, 
$\sigma(t)$ still weaves around $\sigma=1/3$, which means that 
the simulation-time dependence (\ref{se-dis}) at $T_{\rm c}$ with 
$\sigma \approx 1/3$ is optimal and that small discrepancy 
from it can be absorbed into that formula.

\subsection{Three-dimensional Ising model around $T_{\rm c}$}
\begin{figure*}
\includegraphics[width=59mm]{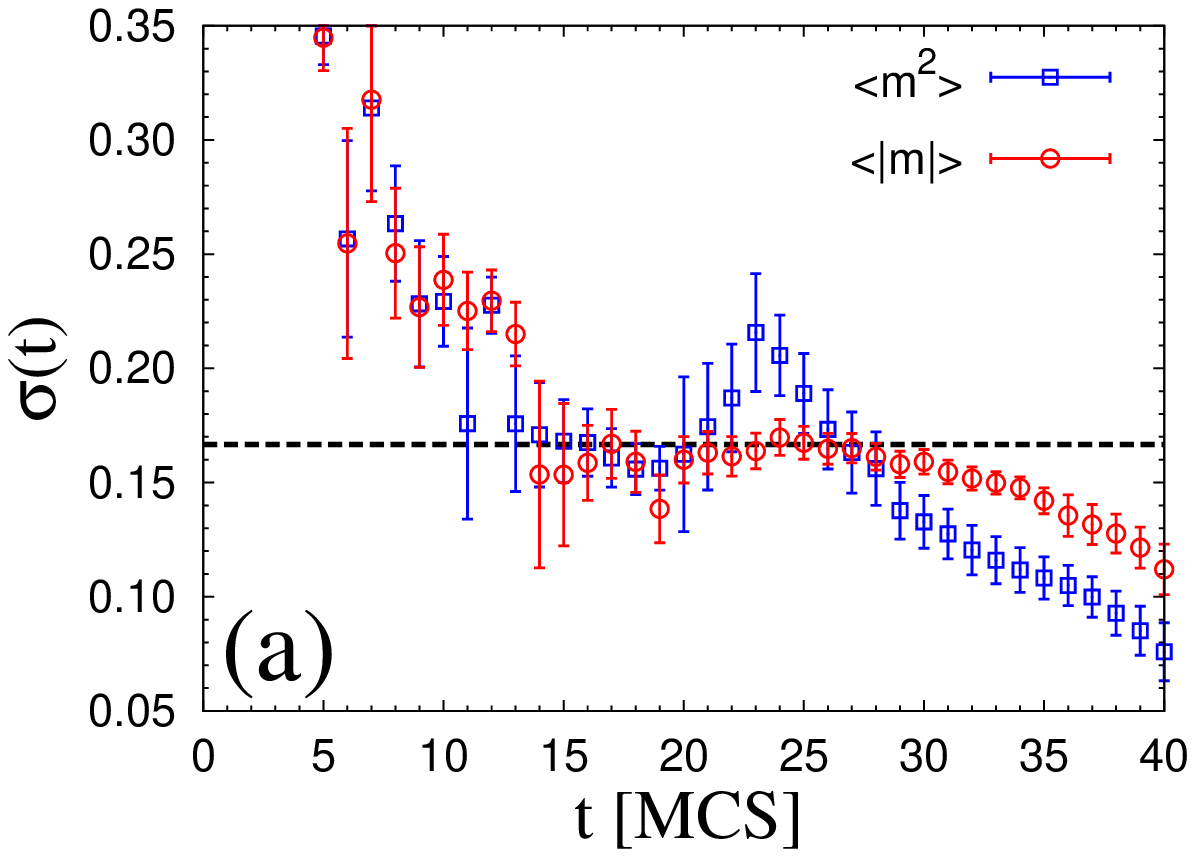}
\includegraphics[width=59mm]{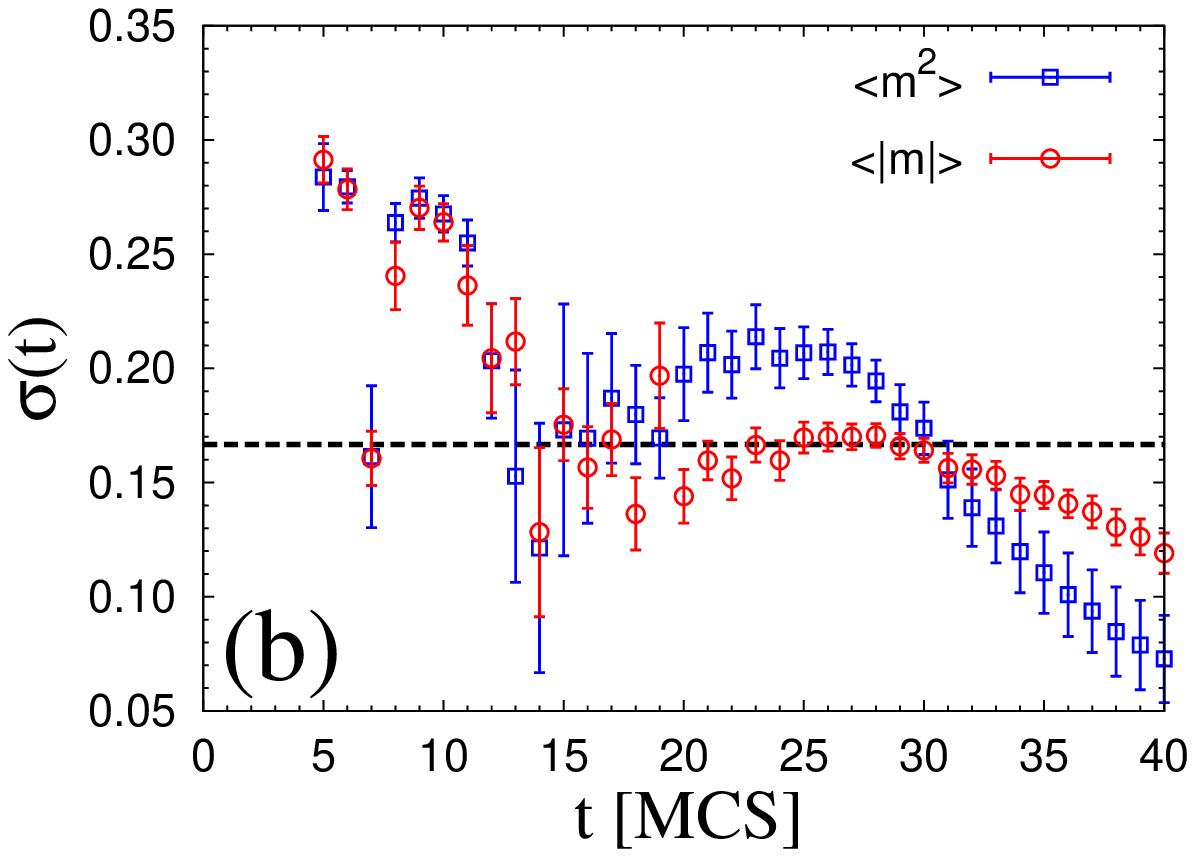}
\includegraphics[width=59mm]{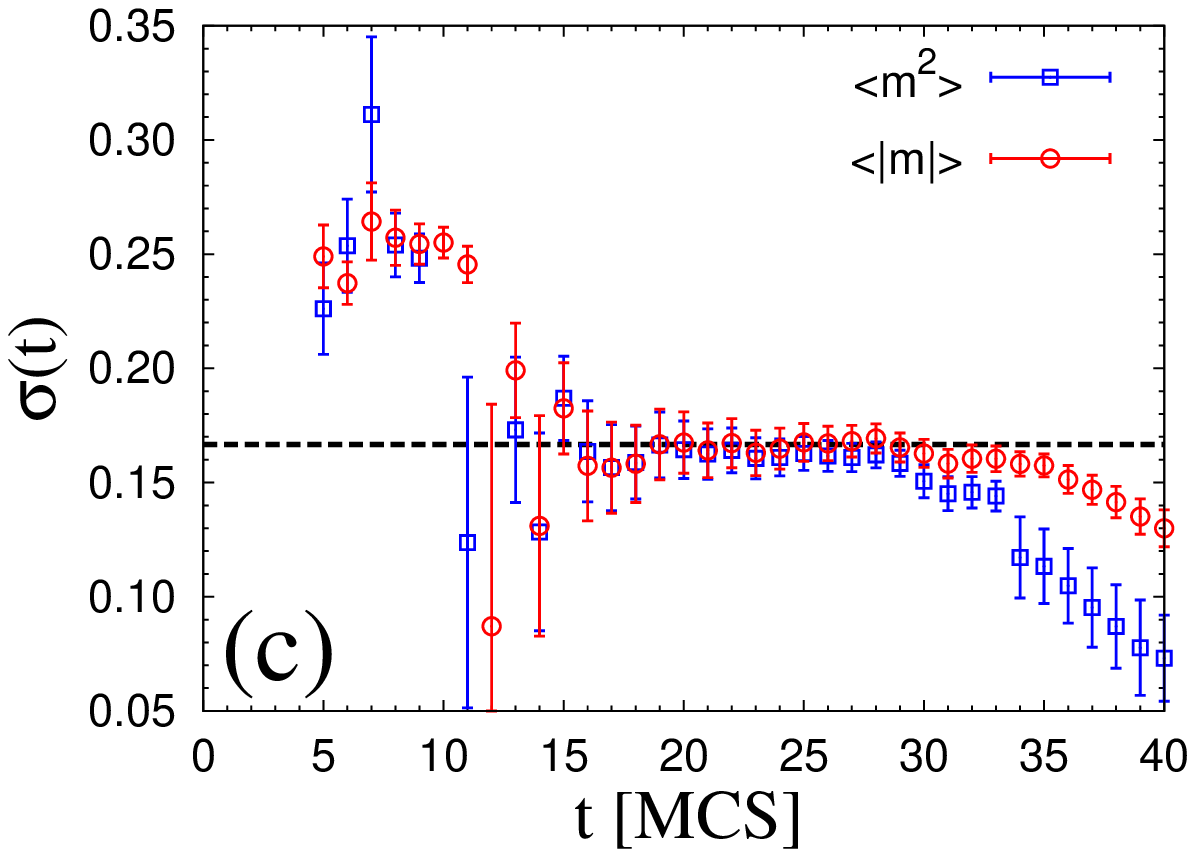}
\includegraphics[width=59mm]{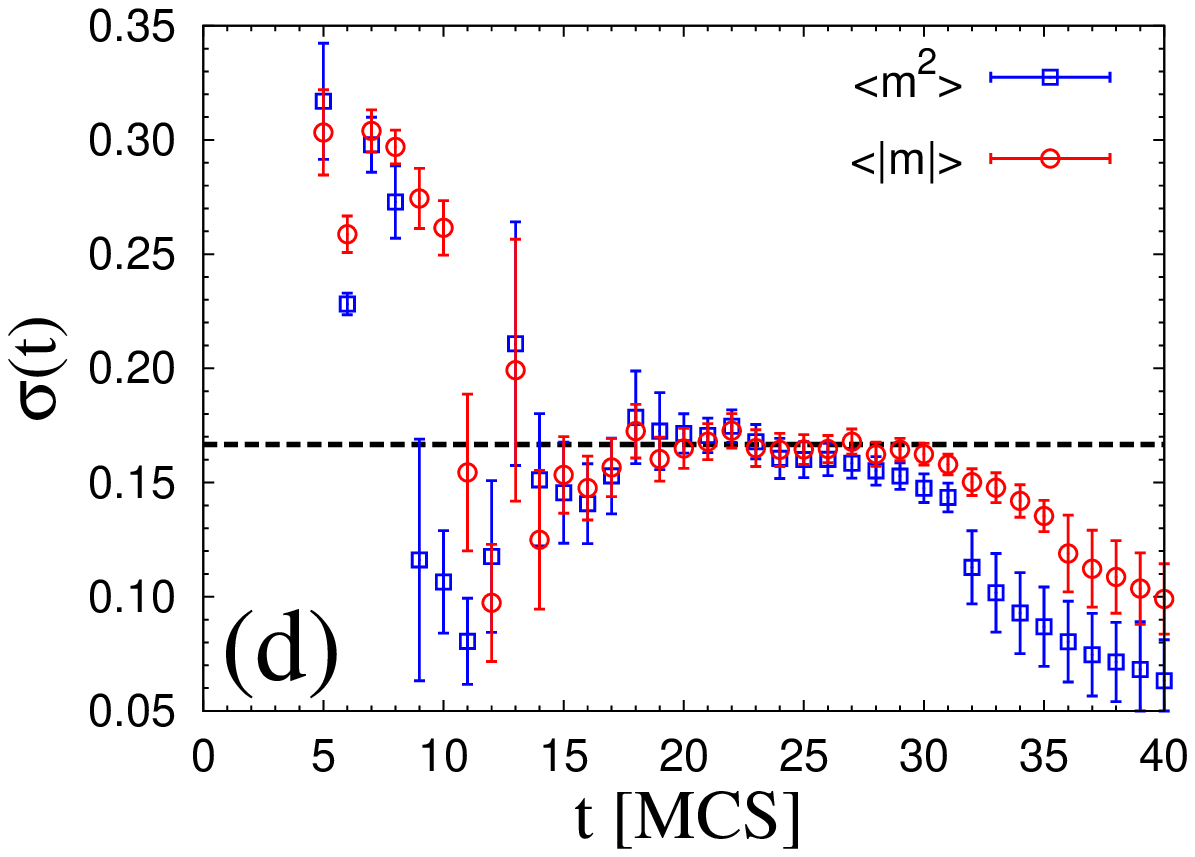}
\includegraphics[width=59mm]{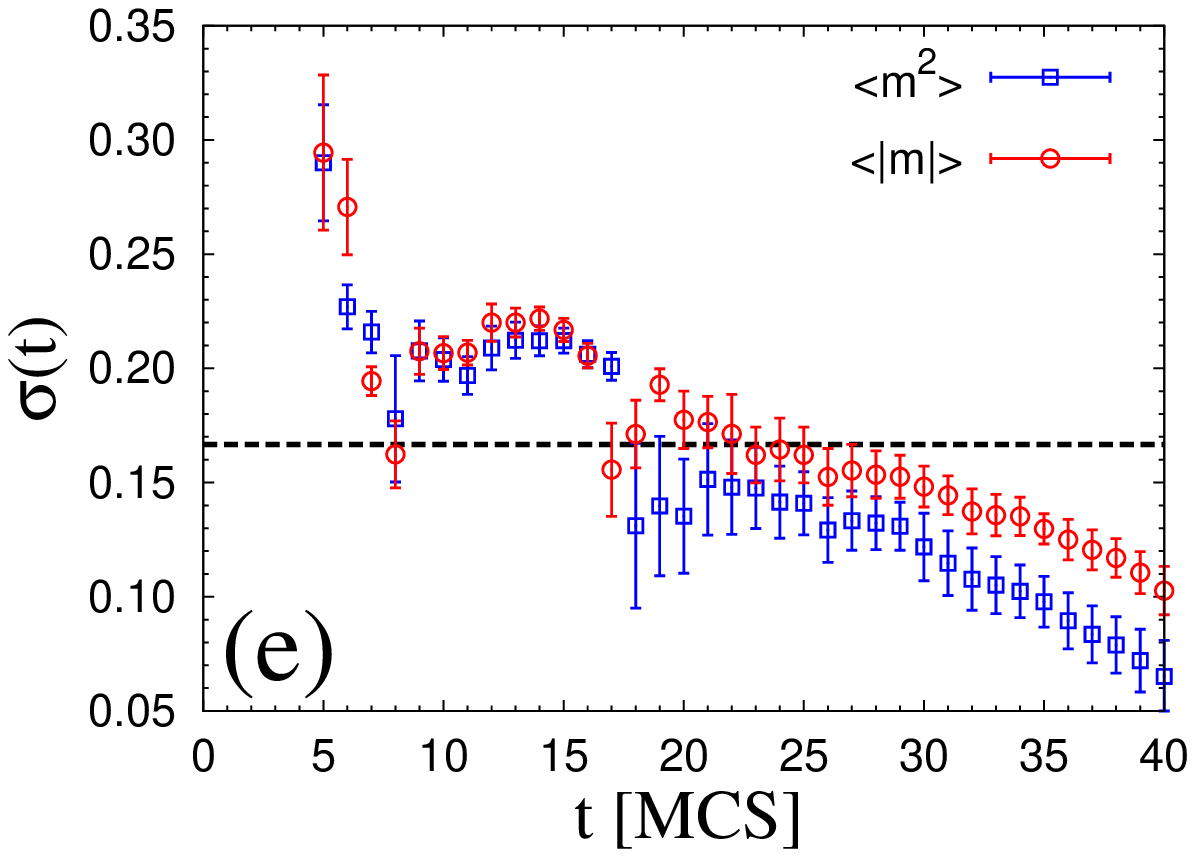}
\includegraphics[width=59mm]{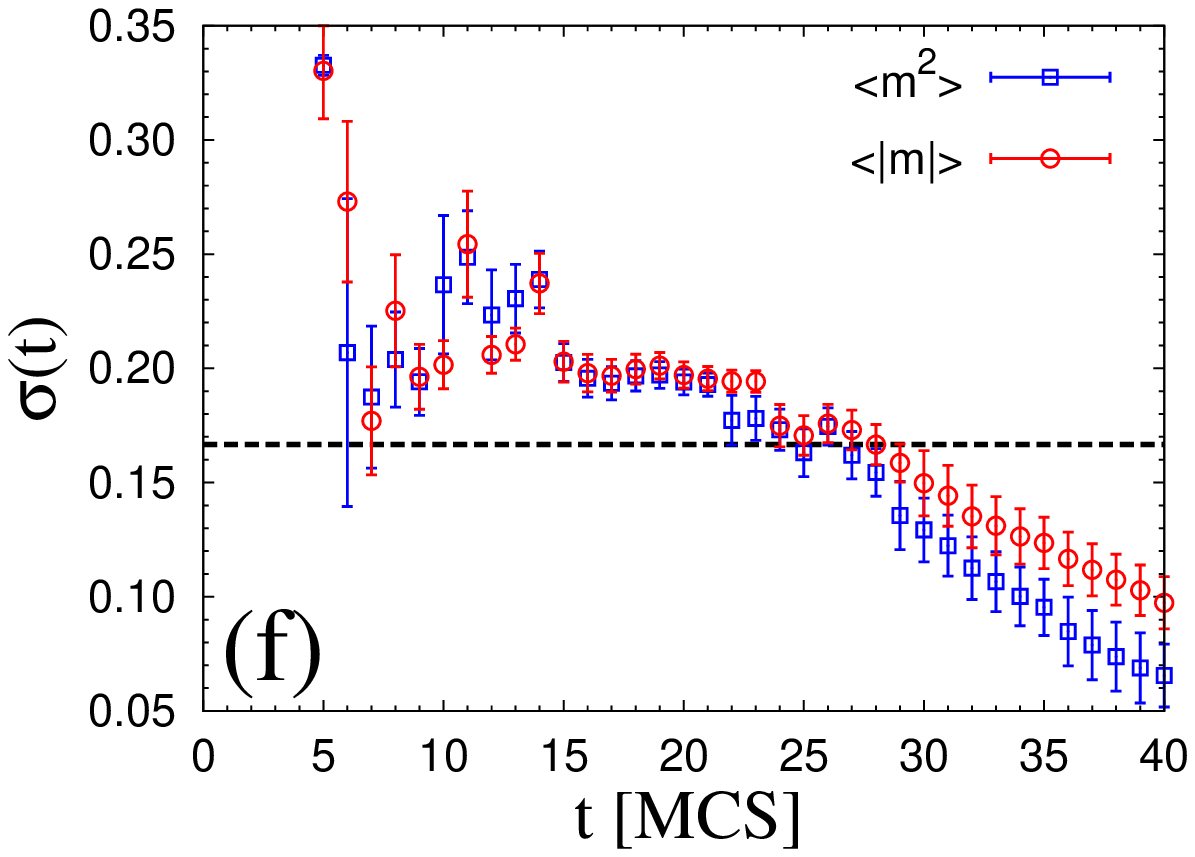}
\caption{(Color online) Relaxation exponent $\sigma(t)$ evaluated 
from the least-squares fitting of the data from $t_{0}(t)$- to $t$-MCS 
($t_{0}(t)$ is determined so as to minimize the residue of the fitting) of 
$\langle m^{2}(t) \rangle$ (squares) and $\langle |m(t)| \rangle$ (circles) 
for $L=560$ at (a) $T=4.511514 J/k_{\rm B}$, (b) $4.511518 J/k_{\rm B}$, 
(c) $4.511520 J/k_{\rm B}$, (d) $4.511522 J/k_{\rm B}$, 
(e) $4.511524 J/k_{\rm B}$ and (f) $4.511528 J/k_{\rm B}$ 
in the three-dimensional Ising model based on the logarithmic fitting 
formulas~(\ref{m2log}) and (\ref{maglog}) versus $t$ used for the fitting. 
The dashed line corresponds to $\sigma=1/6$ as a guide for eyes.
}
\label{sig3D}
\end{figure*}
Then, similar analysis is also possible in the three-dimensional Ising model, 
even though maximum linear size is smaller and statistical and systematic 
errors are larger than that in the two-dimensional one. 
Moreover, in order to understand nontrivial temperature dependence of 
$\sigma(T)$ displayed in Fig.~\ref{3Dexp}, we should consider both 
$\langle |m| \rangle$ and $\langle m^{2} \rangle$ which consist of 
$B_{2,1} \equiv \langle m^{2} \rangle / \langle |m| \rangle^{2}$.

Although the fitting of $\langle |m| \rangle$ in three dimensions is still possible, 
that of $\langle m^{2} \rangle$ is difficult. Since this quantity is nothing but 
the magnetic susceptibility, its fluctuating behavior around $T_{\rm c}$ is 
beyond the accuracy of the present data. Then, instead of the fitting 
based on Eqs.~(\ref{se-dis}) and (\ref{se-m2}), we take logarithm of 
both sides of them,
\begin{eqnarray}
\label{maglog}
\log \langle |m(t,L)| \rangle     &=&  c_{m} t ^{\sigma} + A_{m},\\
\log \langle m^{2}(t,L) \rangle &=& c_{m^{2}} t ^{\sigma} + A_{m^{2}}.
\label{m2log}
\end{eqnarray}
After these trivial transformations, all the data points turn to contribute 
to the fitting equally, and the process of fitting becomes more stable. 
Then, the fitting of $\log \langle m^{2} \rangle$ becomes possible. 
On the other hand, the data in the initial several MCS including 
larger statistical errors tend to contribute 
too much in this logarithmic fitting, and such data should be 
eliminated. Here we take a systematic approach to delete the 
early-time data one by one until the residue of the fitting is minimized. 
In addition, we take a larger system size $L=560$ ($2.0 \times 10^{4}$ 
RNS are averaged) for this analysis, because data points included 
in the Gaussian region at $T_{\rm c}$ increases as the system size 
increases, as shown in Figs.~\ref{br-2D}(a) and \ref{Gaussian}(b).

In Figs.~\ref{sig3D}(a)-(f), relaxation exponent $\sigma(t)$ 
estimated with the above procedure (each data point is 
obtained from the data from $t_{0}(t)$- to $t$-MCS, 
where $t_{0}(t)$ gives the minimum residue) is plotted 
versus $t$ based on Eqs.~(\ref{maglog}) (circles) and 
(\ref{m2log}) (squares) at (a) $T=4.511514 J/k_{\rm B}$, 
(b) $4.511518 J/k_{\rm B}$, (c) $4.511520 J/k_{\rm B}$, 
(d) $4.511522 J/k_{\rm B}$, (e) $4.511524 J/k_{\rm B}$ 
and (f) $4.511528 J/k_{\rm B}$ with the dashed line 
corresponding to $\sigma=1/6$. Each pair covers the 
regions II, III and IV in Fig.~\ref{3Dexp}, respectively. 
Apparently, behavior of $\sigma$ obtained from 
$\langle |m| \rangle$ in Figs.~\ref{sig3D}(a)-(d) is similar 
to that in Fig.~\ref{sig2D}, which indicates that the regions 
II and III are included in the critical one. The exponent 
obtained from $\langle m^{2} \rangle$ behaves similarly 
to that from $\langle |m| \rangle$ in the region III, while 
the former exceeds the latter in the region II. Although 
these behaviors may not seem consistent with that in 
Fig.~\ref{3Dexp}, it actually is. When almost similar 
functions are divided with each other, the quotient 
may have little parameter dependence and therefore 
it may be too sensitive for a small change of conditions. 
Fictitious rapid change of $\sigma(T)$ in the region III 
in Fig.~\ref{3Dexp} can be explained with this mechanism. 
On the other hand, discrepancy of $\sigma(t)$ in the 
region II rather stabilizes the fitting of $B_{2,1}(t,L)$ 
in this region and results in small change of $\sigma(T)$ 
in Fig.~\ref{3Dexp}. The estimate $\sigma = 0.178(2)$ 
based on $B_{2,1}(t,L)$ in this region is slightly larger 
than $\sigma \approx 1/6$, which is consistent with a 
larger estimate of $\sigma(t)$ from $\langle m^{2} \rangle$ 
than that from $\langle |m| \rangle$ there. 
In the region IV, $\sigma(t)$ from $\langle |m| \rangle$ 
monotonically increases as $t$ decreases, which means 
that the critical relaxation characterized by a specific value 
of $\sigma$ is not observed in this region, even though the 
stretched-exponential relaxation formula still looks plausible 
there. The exponent $\sigma(t)$ from $\langle m^{2} \rangle$ also 
behaves similarly but smaller than that from $\langle |m| \rangle$ 
for any $t$, which results in $\sigma = 0.107(4)$ based on 
$B_{2,1}(t,L)$, which is fairly smaller than $\sigma \approx 1/6$. 
\section{Summary and discussion}
In the present article we analyze the early-time critical relaxation of 
the $(2,1)$-Binder ratio $B_{2,1}$ in the two- and three-dimensional 
Ising models simulated with the Swendsen-Wang algorithm. 
In addition to the well-known size independence in equilibrium, this quantity 
also shows size independence at the onset of relaxation when simulations 
are started from the perfectly-disordered state in the Ising models. 
Recently a size-independent quantity $\xi(t,L)/L$ was shown to be scaled 
by $t^{\sigma}-\ln L^{\rho}$ with the stretched-exponential critical relaxation 
exponent $\sigma$ and the supplemental exponent $\rho$ related with 
the coefficient of power in the stretched-exponential function. 
When $B_{2,1}(t,L)$ is assumed to be scaled with the same 
scaling quantity $t^{\sigma}-\ln L^{\rho}$ at the exact critical 
temperature in the two-dimensional Ising model, we have 
$\sigma=0.3277(1)$ and $\rho=0.3168(3)$, which is consistent with 
the expected value $\sigma=1/3$ and satisfies $\sigma \approx \rho$. 
In the similar analysis in the three-dimensional Ising model, the 
critical temperature is identified with the region where $\sigma(T)$ 
seems to change rapidly as $T_{\rm c}=4.511521(2) J/k_{\rm B}$, 
which is consistent with previous numerical studies. The relaxation 
exponent almost takes a constant value $\sigma = 0.178(2)$ just below 
$T_{\rm c}$, which is slightly larger than $\sigma \approx \rho \approx 1/6$ 
directly evaluated from the critical relaxation formula of magnetization.

In two dimensions, we also analyze the distribution function 
of magnetization $P(|m|)$ at the exact critical temperature. 
In equilibrium and at the simulation time with a flat distribution, 
$P(|m|,L)L^{-\beta/\nu}$ is scaled with $|m|L^{\beta/\nu}$, 
which is consistent with the finite-size scaling of the critical 
magnetization, $m_{\rm c}(L) \sim L^{-\beta/\nu}$. 
In the early-time relaxation, the distribution function 
has the Gaussian form, 
$P(|m|;t,L) \sim \exp \left[ - (|m|/m_{\rm G}(t,L) )^{2} \right]$ 
with $m_{\rm G}(t,L) \sim L^{-d/2} \exp (C t^{\sigma})$. 
This behavior is consistent with the Gaussian onset value 
$B_{2,1}(t=0) = \pi/2$, which is almost unchanged until the 
Gaussian distribution breaks down. The exponent $\sigma$ 
is directly related with simulation-time dependence of the 
width of the Gaussian distribution, and its size dependence 
is proportional to $L^{-d/2}$, which shrinks much faster than 
that of the equilibrium distribution.

In addition, the exponent $\sigma$ is also evaluated directly from 
the stretched-exponential relaxation of magnetization at the critical 
temperature. In two dimensions at the exact $T_{\rm c}$, the residue 
of the fitting based on the stretched-exponential relaxation formula 
takes minimum when all the data within the Gaussian region 
is used, and we have $\sigma=0.330(5)$. 
In three dimensions around the estimated $T_{\rm c}$ from fluctuating 
behavior of $B_{2,1}$, we evaluate $\sigma$ both from $\langle |m| \rangle$ 
and $\langle m^{2} \rangle$ in order to compare with the estimate obtained 
from $B_{2,1} \equiv \langle m^{2} \rangle / \langle |m| \rangle^{2}$. In the 
temperature region identified with $T_{\rm c}$, $\sigma$ obtained from 
the both magnetizations are consistent with $\sigma \approx 1/6$, which 
clarifies that the fluctuating behavior of $\sigma(T)$ in this region is a 
fictitious one owing to too little parameter dependence of $B_{2,1}$. 
In the region just below $T_{\rm c}$, $\langle |m| \rangle$ similarly gives 
$\sigma \approx 1/6$ while $\langle m^{2} \rangle$ larger $\sigma(T)$, 
which results in a slightly larger but stable estimate of $\sigma$. In the 
region just above $T_{\rm c}$, convergent stretched-exponentially 
relaxation in magnetizations cannot be observed anymore, which 
results in a large discrepancy of $\sigma(T)$ from $\sigma \approx 1/6$. 
\bigskip
\par
\section*{Acknowledgments}
The random-number generator MT19937~\cite{MT} was used for 
numerical calculations. Most calculations were  performed on the 
Numerical Materials Simulator at National Institute for Materials 
Science. This study was supported by JSPS KAKENHI 
Grant Number JP16K05493.

\end{document}